\documentclass[aps,prd,twocolumn,nofootinbib,twoside,floatfix,reprint]{revtex4-1}

\usepackage{aas_macros}
\usepackage{natbib}
\usepackage{amsmath}
\usepackage{amssymb}
\usepackage{graphicx}
\usepackage{color}

\newcommand{\p}{\ensuremath{\partial}}

\newcommand{\e}[1]{\ensuremath{{\rm e}^{#1}}}

\newcommand{\der}{\ensuremath{{\rm d}}}
\newcommand{\dir}{\ensuremath{\delta_{\rm D}}}

\newcommand{\eqn}[1]{equation~\eqref{#1}}
\newcommand{\eqns}[1]{equations~\eqref{#1}}

\newcommand{\be}{\begin{equation}}
\newcommand{\ee}{\end{equation}}
\newcommand{\Cal}[1]{\ensuremath{\mathcal{#1}}}

\begin{document}

\title{General relativistic `screening' in cosmological simulations}

\author{Oliver Hahn}
\email{oliver.hahn@oca.eu}
\affiliation{Laboratoire Lagrange, Universit\'e C\^ote d'Azur, Observatoire de la C\^ote d'Azur, CNRS,\\
Blvd de l'Observatoire, CS 34229, 06304 Nice cedex 4, France}

\author{Aseem Paranjape}
\email{aseem@iucaa.in}
\affiliation{Inter-University Centre for Astronomy and Astrophysics,\\ Ganeshkhind, Post Bag 4, Pune 411007, India}

\begin{abstract}
\noindent
We revisit the issue of interpreting the results of large volume cosmological simulations in the context of large scale general relativistic effects. We look for simple modifications to the nonlinear evolution of the gravitational potential $\psi$ that lead on large scales to the correct, fully relativistic description of density perturbations  in the Newtonian gauge. We note that the relativistic constraint equation for $\psi$ can be cast as a diffusion equation, with a diffusion length scale determined by the expansion of the Universe. Exploiting the weak time evolution of $\psi$ in all regimes of interest, this equation can be further accurately approximated as a \emph{Helmholtz} equation, with an effective relativistic `screening' scale $\ell$ related to the Hubble radius. We demonstrate that it is thus possible to carry out N-body simulations in the Newtonian gauge by replacing Poisson's equation with this Helmholtz equation, involving a trivial change in the Green's function kernel. Our results also motivate a simple,  approximate (but very accurate) gauge transformation -- $\delta_{\rm N}(\mathbf{k}) \approx \delta_{\rm sim}(\mathbf{k})\times (k^2+\ell^{-2})/k^2$ -- to convert the density field $\delta_{\rm sim}$ of standard collisionless $N$-body simulations (initialised in the comoving synchronous gauge) into the Newtonian gauge density $\delta_{\rm N}$ at arbitrary times. A similar conversion can also be written in terms of particle positions. Our results can be interpreted in terms of a Jeans stability criterion induced by the expansion of the Universe. The appearance of the screening scale $\ell$ in the evolution of $\psi$, in particular, leads to a natural resolution of the `Jeans swindle' in the presence of super-horizon modes.
\end{abstract}

\maketitle

\section{Introduction}
\label{sec:intro}
\noindent
The paradigm of gravitational instability growing tiny primordial density fluctuations into a complex ``cosmic web'' of large scale structure at late times has had great success in explaining measurements of the anisotropic temperature of the cosmic microwave background (CMB) \cite{Gorski+96,Komatsu+11,Planck15-XIII-cosmoparam} and the spatial distribution of galaxies as seen in large surveys \cite{Huchra+83,Percival+01,Abazajian+09}. While CMB anisotropies are accurately described using linear perturbation theory in general relativity \cite{Dodelson03}, a full appreciation of the nature and nonlinear evolution of the cosmic web requires the use of numerical simulations. The tool of choice for the latter -- the $N$-body method -- routinely uses Newtonian dynamics to follow gravitational instability \cite{Bernardeau+02} (although see \cite{Adamek+13,Adamek+15}). As galaxy surveys start to cover ever-increasing volumes, it has become important to run simulations in very large boxes -- large enough that the size of the box approaches/exceeds the Hubble scale where Newtonian theory is no longer valid. This situation has naturally led to the question of whether or not the results of these large volume Newtonian simulations can be trusted at large scales. 

Several authors have addressed this question \cite{HwangNoh06,ChisariZaldarriaga11,NohHwang12,FlenderSchwarz12,Haugg+12,Bruni+14,Fidler+15}, as well as the closely related (but separate) question concerning the inclusion of large scale general relativistic effects in modelling observable quantities \cite{Yoo+09,BonvinDurrer11,ChallinorLewis11,Bertacca+12,Yoo14}. The understanding that has emerged regarding simulations can be summarized as follows. Simulations of collisionless cold dark matter (CDM) track the evolution of the gravitational potential $\psi$, the fluid peculiar velocity $\mathbf{v}$ and density fluctuation $\delta$ (the latter two using the velocities and positions of a large number of ``particles'') in an expanding background. These fields are initialised at some early time, typically using (i) the results of fully relativistic linear perturbation theory for $\delta$ in the so-called comoving synchronous gauge, (ii) the Zel'dovich approximation for $\mathbf{v}$ and (iii) Poisson's equation for $\psi$. The simulation then updates particle positions using their velocities, the velocities using the Euler equation which involves $\psi$, and $\psi$ itself using Poisson's equation sourced by the density field of the updated particle positions. By a remarkable coincidence, these are \emph{exactly} the equations that must be solved in relativistic linear theory on large scales, \emph{provided} one interprets $\delta$ as being in the comoving synchronous gauge but $\mathbf{v}$ and $\psi$ as being in the \emph{Newtonian} gauge (see Appendix~\ref{app:interpretNbody} for details). This feature of linear theory allows a straightforward interpretation of the results of $N$-body simulations in which $\delta$ was initialised in the comoving synchronous gauge. 
Indeed, \citet{Fidler+15} have shown that one can self-consistently define a new `$N$-body gauge' which retains the Newtonian forms for the continuity, Euler and Poisson equations.
The relativistic calculations of observable quantities can also be explicitly written in this `mixed' or `$N$-body' gauge language \cite{ChallinorLewis11,Yoo14}, meaning that these calculations can in principle be embedded into the output of a large simulation (as one might imagine doing for constructing next-generation mock catalogs).

A couple of issues are worth noting however. 
The above discussion implies that simulations in which the density is initialised in the Newtonian rather than synchronous gauge would be solving the \emph{wrong} equations on large scales, and one can show that this error will lead to spuriously large values of the large-scale potential, which would in turn nonlinearly couple to the small scales and render the entire simulation useless (see, e.g., \cite{Fidler+15} who discuss this issue in the context of a correction to the Zel'dovich approximation; we will also address this point below). Additionally, the above solution to interpreting $N$-body simulations is not very intuitive\footnote{An alternative approach to understanding this problem was presented by \cite{Baldauf+11}, who used the result that the effects of modes with wavelengths longer than the box size can be mapped into an effective background curvature \cite{TormenBertschinger96,Cole97} and advocated simulating only small boxes. This `separate universe' technique has been recently applied in studying nonlinear halo bias \cite{Lazeyras+16,Baldauf+16}, but the large volume simulations typically employed for analysing large scale structure, to the best of our knowledge, continue to use the standard $N$-body method \cite{Crocce+15,Skillman+14,Klypin+14,Heitmann+15}.}; the reason why the large and small scales should not couple strongly is hidden inside a gauge transformation.
 
In this paper we will look for simple, physically intuitive recipes that accurately account for the connection between the evolution of large (linear) and small (nonlinear) scales in the same simulation. We will approach the problem by asking whether a simulation can be initialised and run consistently using the Newtonian gauge for \emph{all} fields. Along the way, we will make simplifying approximations concerning, e.g., the time-evolution of $\psi$; we will demonstrate that these are accurate at better than a few per cent in all regimes of interest. 

In section~\ref{sec:diffusion} we use general relativistic, nonlinear evolution equations to discuss the behaviour of the gravitational potential in an expanding Universe. We use these results to motivate a simple and accurate modification of the standard $N$-body method which clarifies how screening of large scale power arises naturally in the Newtonian gauge due to the expansion of the Universe. The results of our screened simulations are discussed in section~\ref{sec:simulations}. Our results also motivate simple post-processing approximations (both at the level of the density field as well as particle positions) to convert the output of a standard simulation into the Newtonian gauge.
In section~\ref{sec:analytical}, we discuss some interesting physical insights provided by our formulation.
We discuss possible extensions of our work and conclude in section \ref{sec:conclude}. Appendix~\ref{app:interpretNbody} describes the equations behind the $N$-body interpretation described above, Appendix~\ref{app:ICs} describes the methodology for generating the initial conditions used in our simulations, and Appendix~\ref{app:Tmunu} shows how particle positions and velocities in a simulation are mapped to an energy-momentum tensor. 

\section{The evolution of $\psi$}
\label{sec:diffusion}
\noindent
We start by setting up our notation and describing the various regimes we will be interested in, followed by an analysis of the evolution of the gravitational potential $\psi$.

\subsection{Setup}
\label{subsec:setup}
\noindent
We assume a perturbed Friedmann-Lem\^aitre-Robertson-Walker (FLRW) metric described in the conformal Newtonian gauge by \cite{Mukhanov+92}
\be
\der s^2 = a(\tau)^2\left[-\left(1+\frac{2\phi}{c^2}\right)\,c^2\der\tau^2 + \left(1-\frac{2\psi}{c^2}\right)\der\mathbf{x}^2\right]\,,
\label{eq:metric}
\ee
where we set vector and tensor perturbations to zero. The conformal time $\tau$ is related to cosmic time $t$ through $\p_\tau = a\p_t = a\Cal{H}\p_a$, with $\Cal{H}\equiv \p_\tau\ln a = a H$ the conformal or comoving Hubble parameter. The background spatial metric is assumed to be flat, $\der\mathbf{x}^2 = \delta_{ij}\der x^i\der x^j$, $i,j=1,2,3$, and the scale factor $a(\tau)$ satisfies the Friedmann equation
\begin{align}
\Cal{H}^2 &= \frac{8\pi Ga^2}{3c^2}\left(\bar\rho_{\rm m}+\bar\rho_{\rm r}+\bar\rho_{\Lambda}\right)\notag\\
&= H_0^2\left(\Omega_{\rm m}a^{-1} + \Omega_{\rm r}a^{-2} + \Omega_{\Lambda}a^2\right)\,,
\label{eq:background}
\end{align}
where $\bar\rho_X$ is the energy density of component $X$, and $\Omega_{\rm m}+\Omega_{\rm r}+\Omega_{\Lambda}=1$. Throughout, we will be interested in the matter and $\Lambda$-dominated regime and will assume that inhomogeneities are driven by collisionless CDM (pressureless dust). This means that we can ignore the effects of $\Omega_{\rm r}$ (as well as radiation perturbations), and the absence of anisotropic stresses on nearly all scales of interest allows us to set $\phi=\psi$ \cite{MaBertschinger95,Bernardeau+02,Bartolo+07}. Note that, in our convention, the metric potentials have units of velocity squared. We will also neglect the effect of baryons at late times (i.e. $a\gtrsim10^{-2}$) and assume that all matter can be accurately described as a single collisionless fluid, as is commonly done in $N$-body simulations.

CDM inhomogeneities in the conformal Newtonian gauge are described by the density perturbation $\delta$ and peculiar velocity $\mathbf{v}$, which are defined using moments of the phase space distribution function for CDM and can be written in terms of the CDM energy-momentum tensor $T^\mu_{\phantom{\mu}\nu}$ as \cite{MaBertschinger95,Bartolo+07} (see also Appendix~\ref{app:Tmunu})
\begin{align}
T^0_{\phantom{\mu}0} &= -\bar\rho_{\rm m}(1+\delta)\,,\notag\\
T^0_{\phantom{\mu}i} &= \bar\rho_{\rm m}(1+\delta)\,v_i/c\,,\notag\\
T^i_{\phantom{\mu}j} &= \bar\rho_{\rm m}(1+\delta)\,v^i\,v_j/c^2\,,
\label{eq:Tmunu}
\end{align}
where the spatial index on $v^i$ is raised and lowered using the Kronecker delta $\delta_{ij}$. Some comments are in order regarding the assumptions underlying these expressions. We are essentially interested in two different regimes of perturbation, the large scale linear regime and the deep sub-horizon regime of nonlinear $\delta$. In the former we expect the perturbative ordering
\be
\delta \sim \frac{v}{c} \sim \frac{\psi}{c^2} = \Cal{O}(\epsilon)\,,
\label{eq:linpert}
\ee
where $ \epsilon\ll1$ is set by the initial conditions, while in the sub-horizon nonlinear regime we expect \cite{Peebles80}
\begin{align}
\delta &= \Cal{O}(1)\quad;\quad \frac{v}{c} = \Cal{O}(\epsilon)\quad;\quad \frac{\psi}{c^2} = \Cal{O}(\epsilon^2)\,,
\label{eq:nonlinpert}
\end{align}
where $\epsilon = \Cal{H}/(ck) \ll1$ at comoving wavenumber $\mathbf{k}$. In each regime, we will be interested in equations at leading order in the respective $\epsilon$. This justifies the expressions in \eqn{eq:Tmunu} where we discarded terms of order $\psi v/c^3$ and higher, which will never be of importance at this order in either regime. 

Similar considerations tell us that the constraint equation $G^0_{\phantom{\mu}0} = (8\pi G/c^4)T^0_{\phantom{\mu}0}$ (after accounting for the Friedmann equation) can be written as
\begin{align}
&\nabla^2\psi - \frac{3\Cal{H}}{c^2}\p_\tau\psi - \frac{3\Cal{H}^2}{c^2}\psi 
\notag\\
&\phantom{3\Cal{H}^2}
=\frac{4\pi G}{c^2}\,a^2\bar\rho_{\rm m}\delta + \Cal{O}\left(\frac{(\nabla\psi)^2}{c^2},\frac{\Cal{H}^2\psi^2}{c^4}\right)\,,
\label{eq:G00=T00}
\end{align}
where we treat $\p_\tau\psi$ and $\Cal{H}\psi$ on equal footing, and spatial derivatives\footnote{We emphasize that we have ignored terms of order $\sim\psi\nabla^2\psi/c^2$ and $\sim(\nabla\psi\cdot\nabla\psi)/c^2$, both of which we refer to as $\Cal{O}((\nabla\psi)^2/c^2)$. These would contribute to the left hand side of \eqn{eq:G00=T00} by adding the quantity $\left(4\psi\nabla^2\psi + (3/2)\nabla\psi\cdot\nabla\psi\right)/c^2$ at leading order \citep{Bartolo+07,Adamek+13}. Although these terms are clearly negligible in the large scale linear regime of \eqn{eq:linpert}, it might be less obvious why we can ignore these terms in the \emph{nonlinear} regime of \eqn{eq:nonlinpert} when they become comparable to terms involving  $\Cal{H}^2\psi$ and $\Cal{H}\p_\tau\psi$. The reason is that, in this regime, \emph{all} these terms are $\Cal{O}(\epsilon^2)$ corrections to the term involving $\nabla^2\psi$. In the linear regime, however the terms $\sim\Cal{H}^2\psi$ and $\sim\Cal{H}\p_\tau\psi$ are of course relevant. In the transition between these two regimes, on grounds of continuity we expect that the contribution of the terms involving $\psi\nabla^2\psi/c^2$ and $(\nabla\psi)^2/c^2$ continues to remain subordinate in comparison with $\nabla^2\psi$.
Equation~\eqref{eq:G00=T00}, as it stands, therefore gives us a numerically convenient way of tracking various terms during all the regimes in which they become important.} are with respect to comoving coordinates $\mathbf{x}$. 

The nonlinear evolution of $\delta$ and $\mathbf{v}$ can be derived most easily by taking moments of the collisionless Boltzmann equation for CDM \cite{Bernardeau+02,Bartolo+07}. The Euler equation for $\mathbf{v}$ becomes
\begin{align}
&\p_\tau\mathbf{v} + \Cal{H}\mathbf{v} + (\mathbf{v}\cdot\nabla)\,\mathbf{v}\notag\\
&\phantom{\p_\tau\mathbf{v}}
=-\nabla\psi + \Cal{O}\left(\frac{\Cal{H}\psi\mathbf{v}}{c^2},\frac{\Cal{H}v^2\mathbf{v}}{c^2},\frac{\psi\nabla\psi}{c^2},\frac{v^2\nabla\psi}{c^2}\right)\,,
\label{eq:Euler}
\end{align}
where shear generated by shell-crossing can be safely ignored since velocities remain non-relativistic. The continuity equation under similar approximations reads
\begin{align}
&\p_\tau\delta + \nabla\cdot\left[(1+\delta)\mathbf{v}\right]\notag\\
&\phantom{\delta+\nabla\delta}
=3\frac{\p_\tau\psi}{c^2} + \Cal{O}\left(\frac{\mathbf{v}\cdot\nabla\psi}{c^2},\frac{\delta\p_\tau\psi}{c^2}\right)\,,
\label{eq:continuity}
\end{align}
where terms of order $\sim\delta\p_\tau\psi$ can be neglected because they are always much smaller than $\p_\tau\delta$.

\subsection{The diffusion of $\psi$}
\label{subsec:diffusion}
\noindent
The Euler and continuity equations are correctly evolved in simulations, except for the term $\sim3\p_\tau\psi$ in the latter (Appendix~\ref{app:Tmunu}). We will discuss the role of this term later, and focus here on the constraint equation \eqref{eq:G00=T00}, which is easily rewritten as
\be
\frac{3a\Cal{H}^2}{c^2}\,\p_a(a\psi) - \nabla^2(a\psi) =  -\frac32\Omega_{\rm m0}H_0^2\,\delta \equiv S\,,
\label{eq:diffusion-a}
\ee
Defining $u\equiv a\psi$ and introducing the variable $\beta$ such that $\p_\beta = (3a\Cal{H}^2/c^2)\p_a$, this becomes the diffusion equation with a source \cite{Mukhanov+92}:
\begin{equation}
\p_\beta u - \nabla^2 u = S\,.
\label{eq:diffusion}
\end{equation}
Of course, the ``source'' $S$ is coupled to $u$ through its own evolution equation, so this is only part of a nonlinear system. Formally, though, the causal Green's function for \eqn{eq:diffusion} is 
\begin{align}
&G(\beta,\beta^\prime,\mathbf{x},\mathbf{x}^\prime) = \left(4\pi\Delta\beta\right)^{-3/2}\,\e{-r^2/4\Delta\beta}\,\theta(\Delta\beta)\,,
\label{eq:greenfuncdiffusion}
\end{align}
where $\Delta\beta = \beta-\beta^\prime$, $r^2 = \|\mathbf{x}-\mathbf{x}^\prime\|^2$ and $\theta$ is the Heaviside step function. This clearly shows that $\psi$ only responds to $\delta$ over length scales $r\sim\sqrt{\Delta\beta}$. The variable $\beta$ can be expressed in terms of the scale factor using
\be
\beta = \frac{c^2}{3}\int^a\frac{\der a^\prime}{a^\prime\Cal{H}^2(a^\prime)} 
=\int^a\frac{\der a^\prime}{a^\prime} \lambda(a^\prime)^2\,,
\label{eq:betadef}
\ee
where we introduced the length scale
\be
\lambda(a) \equiv c/(\sqrt{3}\Cal{H})\,,
\label{eq:lambdadef}
\ee
which is essentially the comoving Hubble scale, apart from the constant factor of $\sqrt{3}$. 

To try and understand the physical significance of the diffusion scale $\beta$, it is useful to contrast it with the other natural integrated scale in the problem, namely the comoving particle horizon $\ell_{\rm p}$ given by
\be
\ell_{\rm p}(a) \equiv \int_0^t\frac{c\,\der t}{a(t)} =c\,\tau 
= \sqrt{3}\int_0^a\frac{\der a^\prime}{a^\prime} \lambda(a^\prime)\,.
\label{eq:lpdef}
\ee
Although $\beta$ and $\ell_{\rm p}$ are related (e.g., for an Einstein-deSitter universe, $\lambda\propto\sqrt{a}$ so that $\beta=\lambda^2=\ell_{\rm p}^2/12$), they are conceptually quite different. Whereas $\ell_{\rm p}$ follows from the finite speed of light and is well-defined even if $a(t)={\rm constant}$, $\beta$ only makes sense in an expanding universe. On scales $k\gg\beta^{-1/2}$, one recovers the Poisson equation, and on scales $k\ll\beta^{-1/2}$, the potential does not respond to the source term, and instead initial differences $S+\nabla^2 u$ are frozen in. For either $c\to\infty$ or $\Cal{H}\to0$ one finds $\beta\to\infty$ so that the Poisson equation is always valid in those limits.

The damping of the response of $\psi$ to $\delta$ at scales larger than $\sim\sqrt{\beta}$ is therefore reminiscent of a Jeans stability criterion -- small scale pressure has now been replaced with a large scale expansion, and only \emph{small} enough wavelengths can gravitate, with $\sqrt{\beta}$ acting like a Jeans length. Although the effect above is gauge dependent (e.g., it disappears in the mixed gauge setup described in Appendix~\ref{app:interpretNbody}), it is interesting that the conformal Newtonian gauge explicitly brings out this connection with an expansion-related Jeans length. We will explore some further connections with Jeans stability analysis below. 

The previous discussion implies that evolving the Newtonian gauge density $\delta$ in a simulation requires solving the diffusion-like \eqn{eq:diffusion} for $\psi$ rather than Poisson's equation. While numerically tractable, it would be far more practical to have a simple modification of standard $N$-body codes or, better yet, a post-processing approximation for standard $N$-body \emph{outputs}, that can accurately reproduce the large scale behaviour of $\delta$ in the Newtonian gauge. We explore this below.

\subsection{A Helmholtz equation for $\psi$}
\label{subsec:helmholtz}
\noindent
The key assumption we will make is that $\psi$ evolves weakly with time in \emph{all} the regimes of cosmological interest. This is true at all scales in the linear regime, where $\psi$ is frozen to be a constant during matter domination and decays slowly when the cosmological constant dominates, and it is also true in the nonlinear regime of $\delta$ due to the non-relativistic nature of CDM velocities (see equation~\ref{eq:nonlinpert}). We therefore introduce the ansatz 
\be
\psi(\tau,\mathbf{x}) \approx \Psi(\mathbf{x})\,D_1(a)/a +\,\textrm{corrections,}
\label{eq:psiansatz}
\ee
where $D_1(a)$ is the \emph{linear theory growth factor for} $\delta$ and we assume that the `corrections' are small. Note that the corrections vanish in linear theory once the growing mode dominates, as can be easily checked by simultaneously solving the Euler equation and the second of equations \eqref{eq:linearisedconstraint}. 
This allows us to write $\p_\tau\psi\approx\psi\p_\tau\ln(D_1/a)=\Cal{H}(f-1)\psi$ where $f\equiv\der\ln D_1/\der\ln a$. The constraint \eqn{eq:G00=T00} then becomes
\be
\nabla^{2}\psi -\ell^{-2} \psi = (3/2)\Omega_{\rm m}H_0^2\,\left(\delta/a\right)\,,
\label{eq:helmholtz}
\ee
where we defined
\be
\ell\equiv\lambda/\sqrt{f}
\label{eq:elldef}
\ee
and $\lambda(a)$ was defined in \eqn{eq:lambdadef}. 

It should not be surprising that an ansatz such as \eqn{eq:psiansatz} is necessary if we decide not to solve the correct diffusion \eqn{eq:diffusion} for $\psi$; the information we neglect by doing so needs to be accounted for. Our ansatz above essentially says that we can use our knowledge of linear theory to supplement the fully nonlinear evolution of the fields in the Newtonian gauge. We will see later that this information can also be brought back using a time-dependent correction to particle positions in standard $N$-body simulations (see also \cite{ChisariZaldarriaga11}). It is clear, however, that this approximation requires (phase-) velocities to be non-relativistic, since changes in matter perturbations need to be slow compared to the diffusion time scale of the potential.

Equation~\eqref{eq:helmholtz} is a Helmholtz equation; its Green's function is given by $G_{\rm H}(r) = \e{-r/\ell(a)}/(4\pi r)$, or 
\be
G_{\rm H}(k) = -1/(k^2+\ell^{-2})
\label{eq:GreenHelmholtz}
\ee
in Fourier space. Comparing this with the Green's function for Poisson's equation, $G_{\rm P}(r)=1/(4\pi r)$ or 
\be
G_{\rm P}(k) = -1/k^2\,, 
\label{eq:GreenPoisson}
\ee
we see the role of the `screening' scale $\ell$ which regulates the small $k$ divergence of $G_{\rm P}$ (equivalently, it introduces an exponential damping in real space at separations larger than $\ell$). The scale $\ell$ is -- as expected -- closely related to the Hubble scale. In fact, during matter domination we have $f=1$ so that $\ell=\lambda$, while $f$ decreases slowly at late times, approximately as $f\approx\left(\Omega_{\rm m}(a)\right)^{4/7}$ \cite{Lahav+91}, so that $\ell$ is very accurately given (to within $\sim2$ per cent at $a=1$) by
\be
\ell(a) = \ell_0\,a^{2/7} \left(\Cal{H}/H_0\right)^{-3/7}\,,
\label{eq:ellapprox}
\ee
where $\ell_0\equiv c/(\sqrt{3}H_0)\,\Omega_{\rm m}^{-2/7}$ and $\Cal{H}/H_0$ can be read off from \eqn{eq:background}. 
We note that a similar result was obtained recently by \citet{Eingorn15} using somewhat different approximations, also leading to a Helmholtz equation but with a screening scale $\ell(a)$ that differs from our expression \eqref{eq:elldef}.

\begin{figure}[t]
\includegraphics[width=0.47\textwidth]{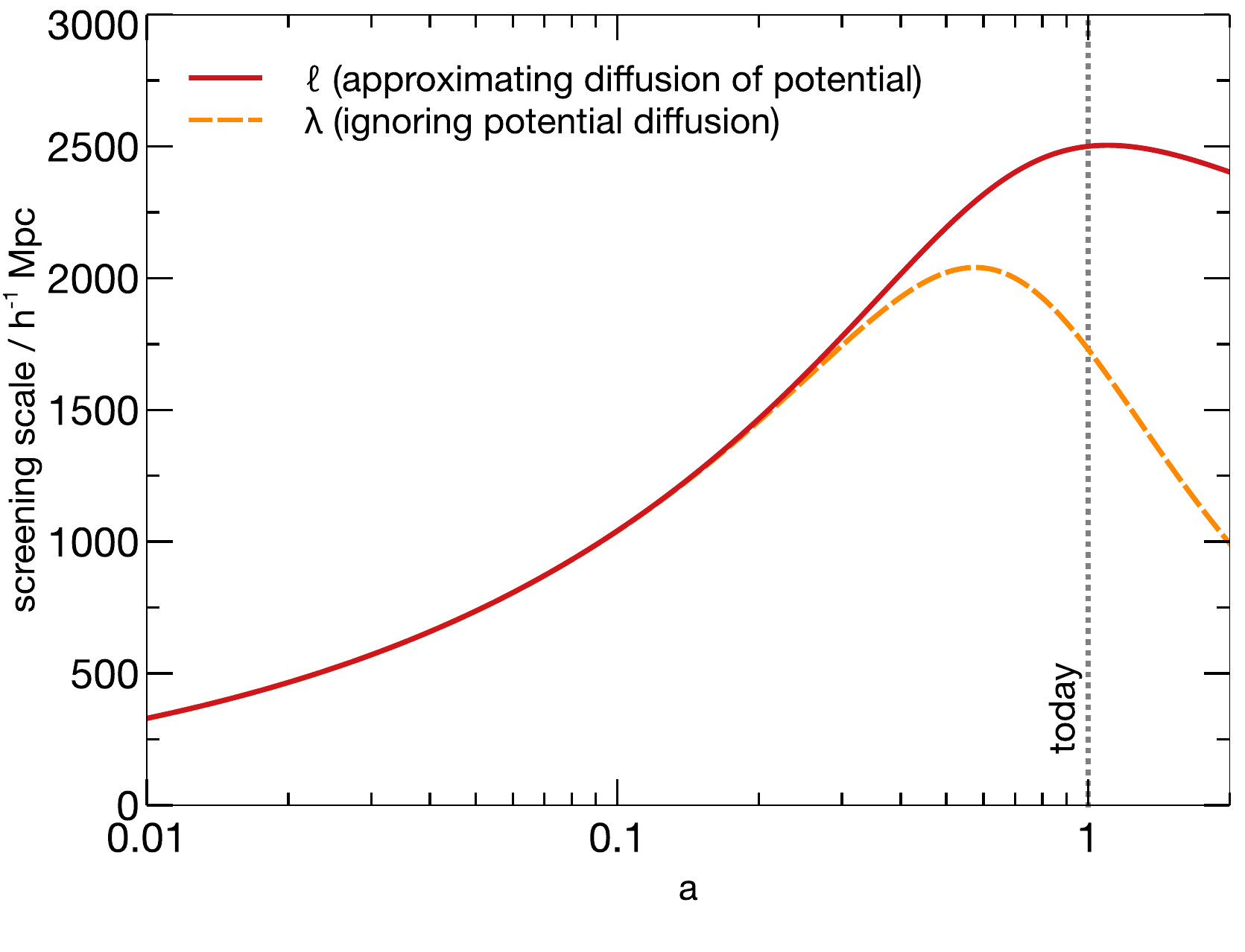}
\caption{\label{fig:screening_scale}The comoving screening scales $\ell(a)$ (solid red line) and $\lambda(a)$ (dashed orange line) in a $\Lambda$CDM universe. Note the decrease at late times due to the cosmological constant.}
\end{figure}

\begin{figure}[t]
\includegraphics[width=0.48\textwidth]{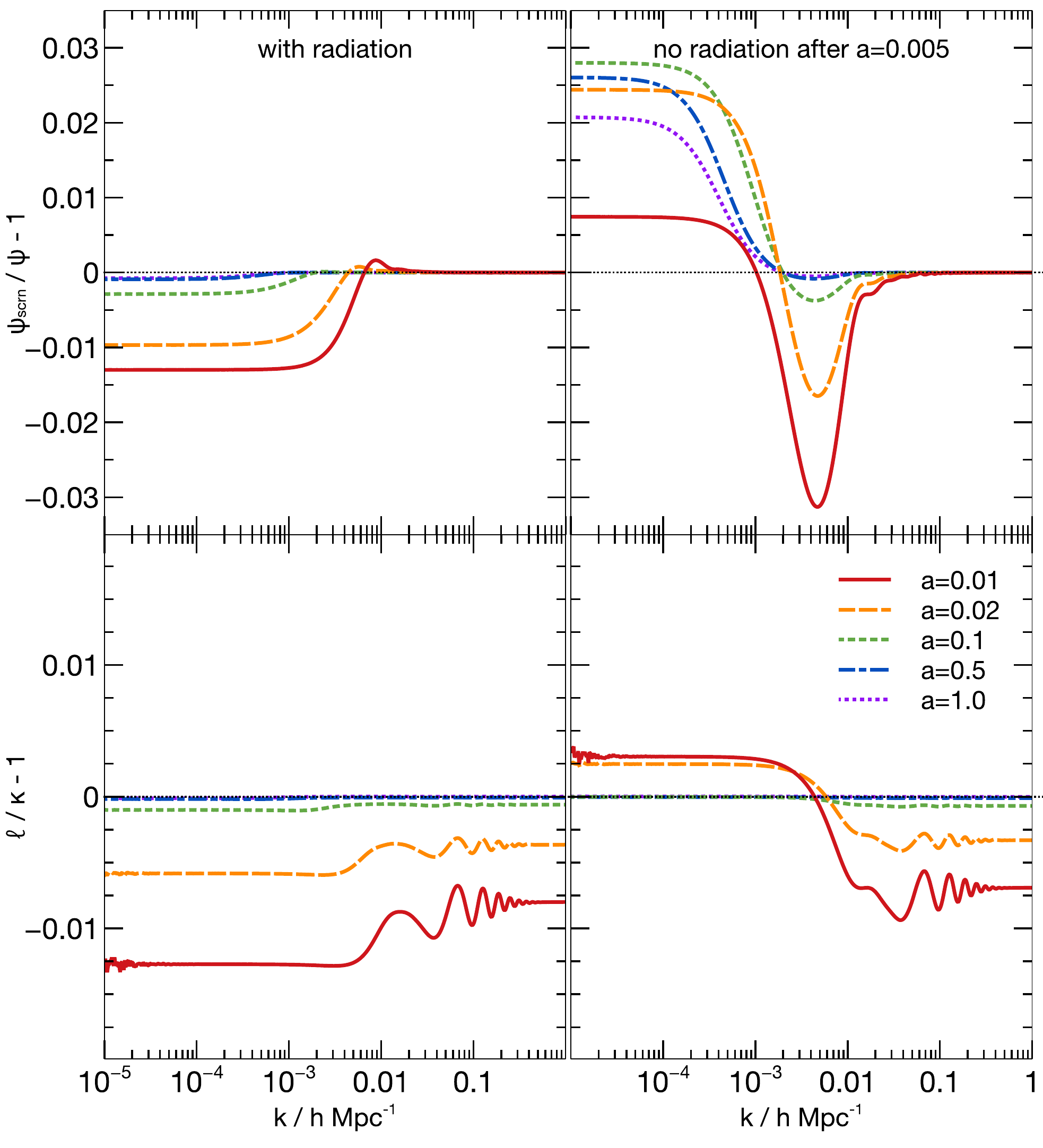}
\caption{\label{fig:ratio_potentials}\emph{(Top row:)} Relative difference between the approximate screened potential $\psi_{\rm scrn}$ obtained from \eqn{eq:helmholtz} sourced by the linear theory $\delta$ in the Newtonian gauge to the full linear theory Newtonian gauge solution for $\psi$. 
We see in the \emph{left panel} that the approximation is accurate at better than $1$ per cent at nearly all times. If we neglect radiation for $a>0.005$ as described in the text, the accuracy becomes $\lesssim3$ per cent (\emph{right panel}).
\emph{(Bottom row:)} Relative difference between the screening scale $\ell$ (scale-independent) and the inferred screening scale $\kappa(k)$ from the gauge transformation, \eqn{eq:gauge_trafo}, between the synchronous and the Newtonian gauge. 
This approximation is accurate at better than $1$ per cent at nearly all times, both with (\emph{left panel}) and without radiation (\emph{right panel}).}
\end{figure}

Figure~\ref{fig:screening_scale} shows the evolution of the screening scale $\ell(a)$ in our fiducial $\Lambda$CDM cosmology\footnote{While we have plotted the exact result for $\ell(a)$ in terms of the growth factor, we note that the expression in \eqn{eq:ellapprox} would be nearly indistinguishable.}. For comparison, we also show the scaled Hubble radius $\lambda(a)$. Curiously, the screening scale $\ell$ has its maximum exactly around the present epoch. This aggravates to some extent the coincidence problem by making a unique point in the evolution of $\ell(a)$ coincident with $z=0$. 
As a sanity check, the top row in Figure~\ref{fig:ratio_potentials} compares the approximation for $\psi$ arising from \eqn{eq:helmholtz} sourced by the linear theory $\delta$ in the Newtonian gauge with the full linear theory solution for $\psi$ in the Newtonian gauge. 
As shown in the left panel, the approximation is accurate at the sub-percent level at nearly all times of interest\footnote{We note that the two potentials $\phi$ and $\psi$ are different at the few per cent level at the earliest times we show, due to radiation shear. Also, note that our approximation uses the linear theory growth function $D_1(a)$ which is only defined after radiation is completely subdominant. As such, we should expect the approximation to work most accurately at fairly late times, and we see that this is indeed the case.}. For our simulations below, we will use a slightly modified background expansion in which we neglect radiation after $a=0.005$ by setting $\Omega_{\rm r}$ suddenly to zero. The right panel show that, even in this case, the discrepancy is $\lesssim3$ per cent at all times.

Finally, we note that switching from Poisson's equation in standard $N$-body codes to the Helmholtz equation \eqref{eq:helmholtz} requires the trivial replacement $k^{-2}\to\left(k^2+\ell^{-2}\right)^{-1}$ in the Green's function kernel. We present the results of this replacement in section~\ref{sec:simulations} below. 
This correction should also be equivalent to the gauge transformation between the comoving synchronous gauge density $\delta_{\rm syn}$ and the Newtonian gauge density $\delta_{\rm N}$, given by \cite{MaBertschinger95}
\begin{equation}
\delta_{\rm N}-\delta_{\rm syn}=-(3\Cal{H}/k^2)\theta_{\rm N}\,,
\label{eq:gauge_trafo}
\end{equation}
where $\theta\equiv\nabla\cdot\mathbf{v}$ is the velocity divergence (see also equation~\ref{eq:NewtVsSyn}).
In our language, this would be re-cast as $\delta_{\rm N}-\delta_{\rm syn}=(k\kappa)^{-2}\delta_{\rm syn}$, where $\kappa(k,a)$ is \emph{a priori} some time-dependent function of $k$ with dimensions of length. 
We show in the bottom row of Figure~\ref{fig:ratio_potentials} that indeed, as expected, $\kappa\approx\ell$ to better than $1$ per cent at nearly all times, independent of scale, for both the standard calculation (left panel) as well as when ignoring radiation as mentioned above (right panel).

\section{Relativistically screened $N$-body simulations}
\label{sec:simulations}
\noindent
We have performed `screened' $N$-body simulations replacing the Poisson kernel $-1/k^2$ with the Helmholtz kernel $-1/\left(k^2+\ell^{-2}\right)$. We describe the simulations and compare the results for the matter power spectrum with those of standard simulations in what follows.

\noindent
\subsection{Linear theory calculations}
\label{subsec:lineartheory}
\noindent
For the full linear theory calculations, we use a re-implementation of both the Newtonian and the comoving synchronous equations of \citet{MaBertschinger95}, very similar to their original {\sc Linger} code. The same calculation could also be performed using, e.g., the code {\sc class}\footnote{http://class-code.net} \cite{Lesgourgues11-CLASS}. Some care needs to be taken to match the linear calculations (including cold dark matter, baryons and radiation) to the collisionless single-fluid $N$-body simulations discussed below. Since radiation needs to be included at early times in order to produce realistic density spectra, we did include it properly up to $a=0.005$ and explicitly set the radiation density parameter $\Omega_{\rm r}$ to zero afterwards in the linear calculations. When we initialise the $N$-body simulations at $a_{\rm ini}=0.01$ (see our discussion below), residual effects have thus sufficiently decayed. We note however that one could not simply leave $\Omega_{\rm r}\neq0$ in the linear calculation without including it also in the $N$-body code since it affects the linear growth at a small but non-negligible level even after this value of $a_{\rm ini}$. 

\begin{figure*}[t]
\includegraphics[width=0.9\textwidth]{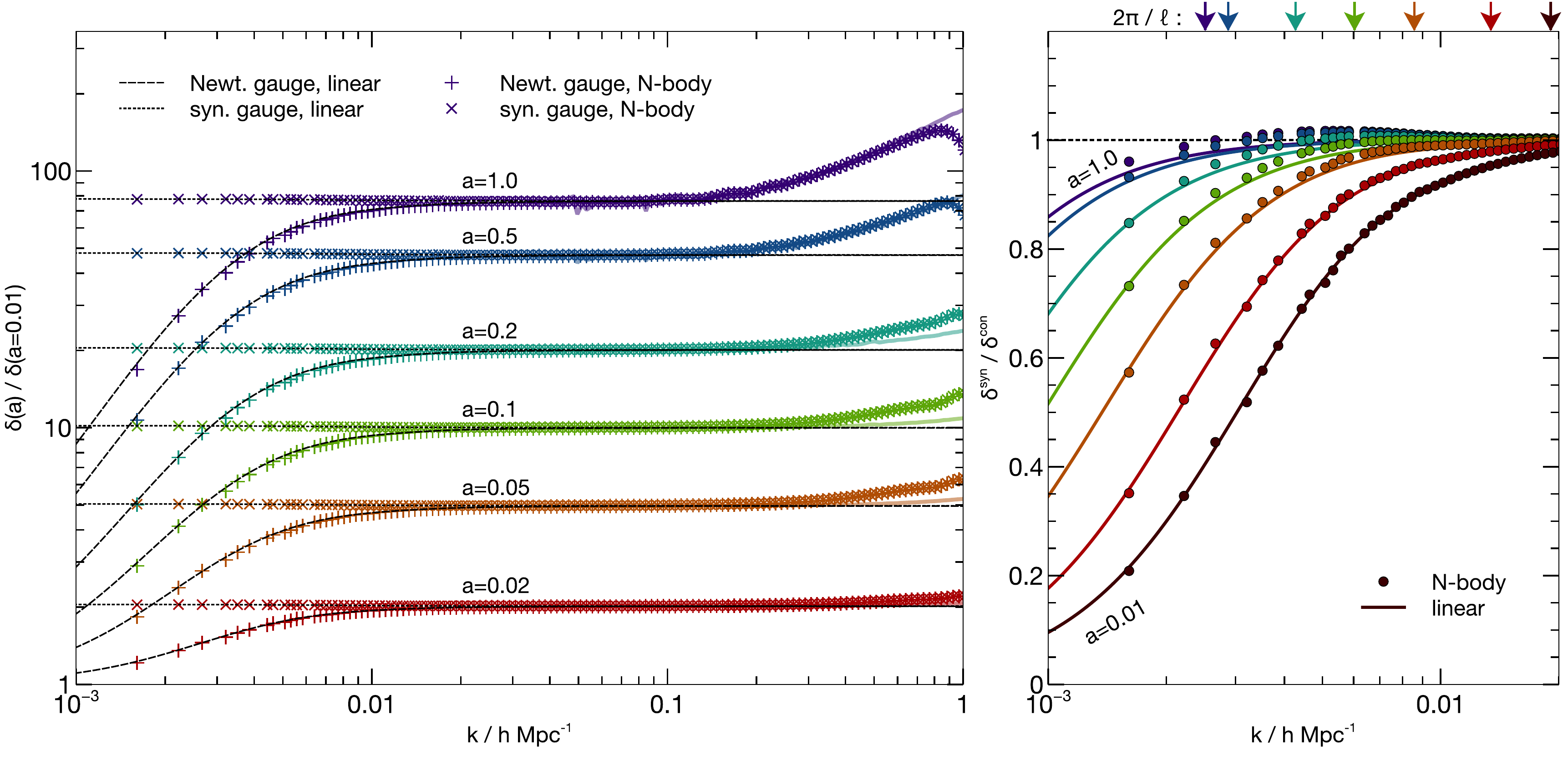}
\caption{\label{fig:Nbody_results} 
\emph{(Left panel:)} The ratio  $\sqrt{P_\delta(k,a)/P_\delta(k,a=0.01)}$, as a function of comoving wavenumber $k$, for several values of scale factor $a$ (increasing from bottom to top). The dotted (dashed) lines show the linear theory result in the comoving synchronous (conformal Newtonian) gauge, while the symbols `$\times$' (`$+$') show the measurements in the standard (screened) simulation. Thick solid lines show the results of smaller simulations to test for resolution effects (see text). \emph{(Right panel:)} The ratio $\sqrt{P_{\delta,{\rm syn}}(k,a)/P_{\delta,{\rm N}}(k,a)}$ of comoving synchronous and Newtonian gauge power spectra from linear theory (solid lines) and the ratio $\sqrt{P_{\delta,{\rm std}}(k,a)/P_{\delta,{\rm scrn}}(k,a)}$ from the standard and screened simulations (filled circles), at scales $k<0.02\,h\,{\rm Mpc}^{-1}$. The lowermost (black) line and circles show the respective ratios at the starting epoch of the simulation $a=0.01$. (Thus, the circles at $a>0.01$ are the ratio of the corresponding `$\times$' and `$+$' from the left panel, multiplied by the lowermost black circles; the latter being set by the initial conditions.) The arrows at the top of the panel mark the wave number corresponding to the screening scale $\ell$ at each epoch, with scale factor increasing from right to left.}
\end{figure*}

\subsection{$N$-body simulations and initial conditions}
\noindent
For the non-linear calculations, we use the tree-PM code {\sc L-Gadget3}  \cite{Angulo2012} to evolve density perturbations and velocities in a simulation box of volume $(4\,h^{-1}{\rm Gpc})^3$ using $1024^3$ $N$-body particles between our initial redshift $z_{\rm ini}=99$ and $z=0$. For all simulations, we assume $\Omega_{\rm m}+\Omega_\Lambda=1$, with a matter density parameter $\Omega_{\rm m}=0.276$, including $\Omega_{\rm b}=0.045$ as the baryon density,  and a cosmological constant consistent with $\Omega_\Lambda=0.724$. We use a Hubble constant $H_0=100h\,{\rm km}/{\rm s}/{\rm Mpc}$ with Hubble parameter $h=0.70$. The large-scale spectral index is $n_s=0.961$ and the power spectrum is normalised so that $\sigma_8=0.811$. Since the $\sigma_8$ normalisation is gauge dependent, we determine the correct $\sigma_8$ for the comoving synchronous spectrum and apply the same normalisation to the Newtonian spectrum, so that the amplitude of sub-horizon density perturbations are identical for every $k$ at the initial time. Given the box size and particle number, our $N$-body particle mass is $m_p\approx1.9\times10^{12}\,h^{-1}{\rm M}_\odot$. We found that, due to the very large box size and the modest particle count we employ, using the tree-force leads to unacceptably large errors on small scales: a well known phenomenon. This is seen in terms of a large drop of power in the density field on small scales before non-linear growth sets in at low redshift. While more elaborate solutions are possible, we have resorted to using only the particle mesh (PM) force in our simulations that is obtained with a PM grid of $2048^3$ cells. This significantly reduces numerical errors. In order to estimate the degree of numerical convergence, we also include results obtained with the same set-up but for a box of $(500\,h^{-1}{\rm Mpc})^3$ using $1024^3$ particles. Overall, we find that results are numerically converged for the purpose of our analysis here.

Initial density and velocity spectra were generated in both Newtonian and comoving synchronous gauge as discussed above. Particle positions and velocities were initialized using the Zel'dovich approximation using the {\sc Music} code \cite{Hahn2011}. The initial conditions for the screened simulations must be set with some care, since these must correspond to linear theory in the Newtonian gauge. In particular, one must include corrections to the Zel'dovich approximation at large scales \cite{ChisariZaldarriaga11}. Fortunately, this is straightforward to implement (and in fact simply corresponds to using the Newtonian gauge transfer functions consistently), and we describe our methodology in detail in Appendix~\ref{app:ICs}. As we show in Appendix~\ref{app:Tmunu}, both the simulations then correctly evolve the particle velocities and positions, up to the correction in the screened simulation due to the term $\sim\p_\tau\psi$ on the right hand side of \eqn{eq:continuity} (we show below that this correction is small). 

\subsection{Power spectrum evolution}
\label{subsec:Nbodyresults}
\noindent
Figure~\ref{fig:Nbody_results} compares the matter power spectra in the screened simulations with those of standard simulations initialised using the same random seed for the potential $\psi$. (Recall that the standard simulations also evolve $\psi$ and $\mathbf{v}$ in the Newtonian gauge, but the density $\delta$ in the comoving synchronous gauge.) The left panel of the Figure shows the quantity $\sqrt{P_\delta(k,a)/P_\delta(k,a=0.01)}$ using linear theory (lines) and the measurements in our simulations (symbols) for both gauges. This ratio is the scale-dependent growth in each gauge, relative to the starting epoch. The linear theory curves demonstrate the well-known facts that (a) large scale modes in the conformal Newtonian gauge grow slower than the corresponding small scale modes and (b) \emph{all} modes in the comoving synchronous gauge grow at approximately the same rate, determined by the growth factor $D_1(a)$ \cite{MaBertschinger95}. The symbols at large scales show that our simulations, both standard and screened, are accurately tracking the respective scale-dependent growth in each gauge.

At small scales, the results from the two simulations become identical and depart from linear theory. To test for resolution effects, we show the results of the smaller simulations described above as the thick solid lines. We see that the results are numerically converged at epochs $a\gtrsim0.2$, with the departure from linear theory being consistent with genuine nonlinear growth. At earlier epochs, we see a resolution and epoch dependent departure at $k\gtrsim0.3\,h\,{\rm Mpc}^{-1}$. 

The level of accuracy with which we reproduce linear theory at large scales is demonstrated in the right panel, which focuses on scales $k<0.02\,h\,{\rm Mpc}^{-1}$ and shows the ratio $\sqrt{P_{\delta,{\rm syn}}(k,a)/P_{\delta,{\rm N}}(k,a)}$ of comoving synchronous and Newtonian gauge power spectra from linear theory as the solid lines, and the ratio $\sqrt{P_{\delta,{\rm std}}(k,a)/P_{\delta,{\rm scrn}}(k,a)}$ from the standard and screened simulations as the filled circles. In addition to the epochs shown in the left panel, the right panel also shows these ratios at the starting epoch of the simulation $a=0.01$. There are two potential sources of error in our screened simulations: (a) we solve the Helmholtz \eqn{eq:helmholtz} instead of the more accurate, diffusion-like \eqn{eq:diffusion}, and (b) as described in Appendix~\ref{app:Tmunu}, the evolution of the density in the screened simulation misses the term $3\p_\tau\psi$ in \eqn{eq:continuity}. The results in the right panel of Figure~\ref{fig:Nbody_results} show that the cumulative effects of these errors are at the few per cent level for all epochs and scales that we explore.

More importantly, the convergence between the standard and screened simulations at small scales, together with the relatively large excess power in the Newtonian gauge as compared to the synchronous at large scales, demonstrates that the small scales in our screened simulations are indeed being correctly screened from the large scale power.

\subsection{A posteriori gauge transformation of $N$-body simulations}
\label{subsec:gaugefix}
\noindent
The simplicity of our modification to the Green's function kernel of the $N$-body simulation ($-1/k^2 \to -1/\left(k^2+\ell^{-2}\right)$) suggests that it should be possible to go further and simply modify the density field of a standard simulation with a \emph{multiplicative factor in Fourier space} and obtain the power spectrum in the Newtonian gauge. Essentially, we expect that the potential is evolved nearly identically in each simulation, motivating the relation 
\be
\delta_{\rm N}(\mathbf{k}) \approx \delta_{\rm sim}(\mathbf{k})\times (k^2+\ell^{-2})/k^2
\label{eq:ppfix}
\ee
as a simple and accurate post-processing approximation to convert the density field $\delta_{\rm sim}$ of a standard simulation into the Newtonian gauge density. 

\begin{figure}[t]
\includegraphics[width=0.45\textwidth]{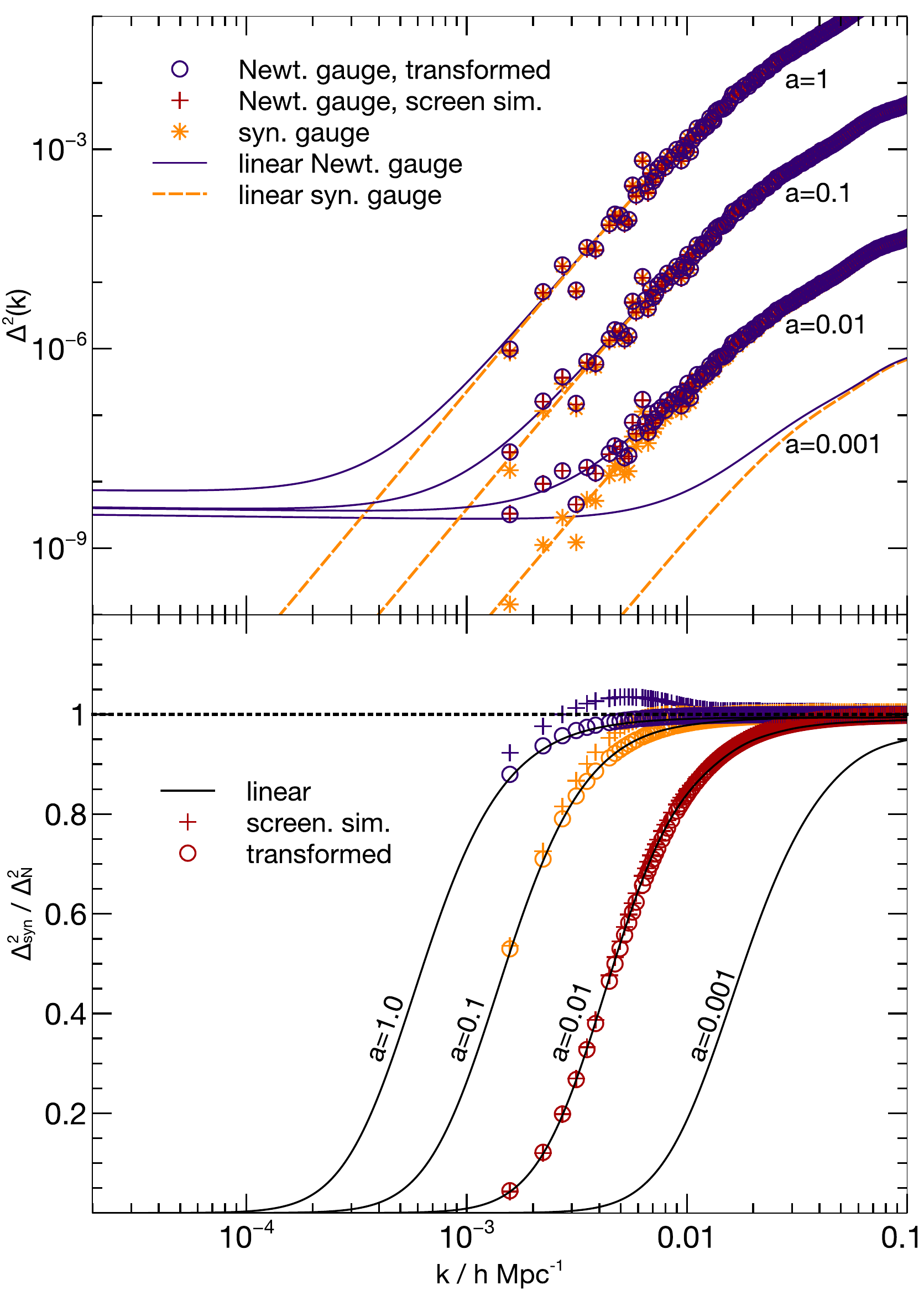}
\caption{\label{fig:quickfix}  
\emph{(Top panel:)} Evolution of the power spectrum in a standard $N$-body simulation of the comoving synchronous gauge density spectrum (orange stars) and of a screened simulation in the Newtonian gauge (red crosses). We also show the power spectra obtained by applying the transformation from \eqn{eq:gaugefix} to convert the synchronous spectrum to the Newtonian gauge by displacing the particle positions (blue circles). The solid lines show the result of the linear theory calculation. The density fluctuations asymptote to a constant on super-horizon scales in Newtonian gauge and to a power-law in synchronous gauge (here $\propto k^{2}$). The amplitude on superhorizon scales in the Newtonian gauge is frozen during matter domination but evolves at earlier times and more importantly due to a cosmological constant.\emph{(Bottom panel:)} Ratio of the power spectrum in comoving synchronous to Newtonian gauge for the spectrum obtained in the Newtonian simulation (crosses) and from the displacement transformed synchronous simulation (circles). Note that the latter provides a more accurate approximation to linear theory at large scales.}
\end{figure}

We can also express this approximated gauge transformation in terms of particle displacements. This is very close in spirit to the calculation presented by \cite{ChisariZaldarriaga11}. Since the effect occurs only on very large scales, the necessary correction can be assumed to be reasonably well described by the Zel'dovich approximation. Then we can write, for sufficiently large scales, the displacement field in the simulation and in the approximated Newtonian gauge as
\begin{equation}
\mathbf{x}_{\rm N} = \mathbf{q} + \nabla \Phi_{\rm N},\quad\textrm{and}\quad\mathbf{x}_{\rm sim} = \mathbf{q} + \nabla \Phi_{\rm sim},\label{eq:zeldo2}
\end{equation}
where $\Phi_{\rm N}$ and $\Phi_{\rm sim}$ are the potentials giving rise to the respective displacement fields (proportional to the velocity potential in standard Lagrangian perturbation theory; see also Appendix~\ref{app:ICs}). Using the approximate relation from eq.~(\ref{eq:ppfix}), we can express the shift between the Newtonian gauge particle positions and the simulation particle positions in Fourier space in terms of a displacement field $\mathbf{L}_{{\rm sim}\to{\rm N}}$ as 
\begin{equation}
\mathbf{L}_{{\rm sim}\to{\rm N}}(\mathbf{k},a) \approx \frac{i\mathbf{k}}{\ell^2(a) k^4}\, \delta_{\rm sim}(\mathbf{k},a), \label{eq:gaugefix}
\end{equation}
where we made the time dependence explicit. We note that the time dependence of the right hand side cancels during matter domination at large scales, where $\delta_{\rm sim}(\mathbf{k},a) \propto a$ and $\ell(a)\propto\Cal{H}^{-1}\propto a^{1/2}$. The scale-dependence is such that, at large scales, the right hand side behaves like $\sim(i\mathbf{k}/k^2)\delta_{\rm N}(\mathbf{k},a_{\rm in})$. 

These results agree with those reported by \citet{ChisariZaldarriaga11} for the matter dominated regime. 
In their formulation, however, the displacement correction is a constant and must only be applied at the initial time. Superficially, it may seem that our results at late times disagree with theirs. As we discuss in Appendix~\ref{app:Tmunu}, however, the difference between our results and those of \cite{ChisariZaldarriaga11} is a consequence of the assumptions we make in converting particle positions into density fields. Briefly, we use an energy-momentum tensor for particles living on an unperturbed grid, but initialised with the relativistic linear theory transfer functions for the density and velocity fields. Consequently, our approach allows us to use the particle positions in a standard simulation to compute the displacement correction at any time and does not require the initial density field of the simulation to be stored. We discuss the accuracy of our prescription below.

The Zel'dovich approximation eq.~(\ref{eq:zeldo2}) requires the gradient to be evaluated at the Lagrangian particle position. Since the particle displacement correction concerns only scales much larger than the evolved (nonlinear) distance between the Eulerian and Lagrangian coordinate for each particle, one can to a good approximation evaluate the gradients simply at the Eulerian particle positions. We use cloud-in-cell interpolation to achieve this. Figure~\ref{fig:quickfix} compares the resulting power spectrum with the one measured in our screened simulations, relative to the standard simulations. We see that our displacement corrected results match the results of linear theory even better than the screened simulations. This is expected for two reasons: firstly, the screened simulations are affected by the accumulation of (sub)percent level errors (see Figure~\ref{fig:ratio_potentials}) over time, and more importantly, these simulations do not account for the term $\sim3\p_\tau\psi$ in \eqn{eq:continuity}. Equation~\eqref{eq:gaugefix} on the other hand uses the synchronous density contrast which obeys the continuity equation \emph{without} this term and therefore correctly combines with the screening scale $\ell(a)$ to reproduce the late time large scale behaviour of $\delta_{\rm N}$. 
\section{Further analytical insights}
\label{sec:analytical}
\noindent

\subsection{Interpreting the screening scale}
\label{subsec:interpret}
\noindent
As we discussed in section~\ref{subsec:diffusion}, the diffusion-like evolution of $\psi$ in \eqn{eq:diffusion} has a natural interpretation in terms of Jeans instability, with an effective Jeans scale $\sim\sqrt{\beta}$ determined by the expansion of the Universe (and therefore related to, but conceptually very different from the particle horizon $\ell_{\rm p}$; compare equations \ref{eq:betadef} and \ref{eq:lpdef}). When the diffusion equation \eqref{eq:diffusion} is approximated as the Helmholtz equation \eqref{eq:helmholtz}, the Jeans length takes on a new form and appears as the relativistic screening scale $\ell$. Although our terminology of `screening' is inspired by the phenomenon of Debye screening in plasma physics \citep[see, e.g.,][]{Sturrock94}, we should note that there are some pitfalls involved with this association.

Debye screening is the phenomenon where the electrostatic potential of a point charge placed at rest in a plasma picks up a multiplicative contribution $\sim\e{-r/\lambda_{\rm D}}$ (where $\lambda_{\rm D}$ is the Debye length) on top of the usual $1/r$, due to the collective behaviour of its neighbouring electrons and ions in the plasma which converts Poisson's equation for the potential into a Helmholtz equation at leading order. The potential of the charge is then said to be `screened' or `shielded' by the plasma. There have been several analyses, in a variety of contexts, discussing the notion of screening in a gravitating system \cite{Marochnik68,Saslaw85,PaddyVasanthi85,Spiegel99}; however, all of these end up with a screening term that has the `wrong' sign in the Helmholtz equation, leading to oscillations rather than exponential damping. Put simply, gravitational effects cannot be shielded.

The fact that our Helmholtz \eqn{eq:helmholtz} \emph{does} have the correct sign to induce damping therefore suggests that the effect we are seeing is a version of Jeans instability (arising from the expansion of the Universe) as discussed above, rather than some form of collective gravitational screening. A further connection with Jeans instability occurs through a natural resolution of the `Jeans swindle' problem that is provided by the Helmholtz \eqn{eq:helmholtz}, as we discuss next.

\subsection{A note on the Jeans swindle}
\label{subsec:jeans}
\noindent
The `Jeans swindle' refers to the problem of making sense of Poisson's equation in the presence of a uniform density \cite{BinneyTremaine87}, or more generally, in the presence of super-horizon modes that are indistinguishable from a uniform density at leading order in spatial derivatives. The problem arises because, in the presence of a uniform density, Poisson's equation is undefined in Fourier space due to its divergence as $k\to0$. The `swindle' comprises of stating that this concerns only the mean potential which gives rise to no gravitational force and thus the problem is -- while mathematically inelegant -- fictitious. More precisely, one ignores the fact that $\nabla^2\phi_0=4\pi G\rho_0$ and $\nabla\phi_0=0$ are inconsistent unless $\rho_0=0$. 

Calculations that account for the expansion of the Universe, on the other hand, have shown that the expansion leads to terms that cancel the divergence if one switches from proper to comoving coordinates \cite{NityanandaPaddy94,Falco+13}. The presence of super-horizon modes, however, renders this solution incomplete, since today's super-horizon modes could be tomorrow's super-clusters, so that cancelling the divergence at each time would require tracking the dynamics of these modes. In other words, super-horizon modes in comoving coordinates pose the same problem as the global mean density did in proper coordinates.

This problem is well known to no longer arise for the Helmholtz equation and, in fact, an exponential weakening of the gravitational potential $\sim\exp(-r/\ell)/r$ was already discussed before the advent of relativistic cosmology \cite{Seeliger1895,Neumann1896} (see also \cite{Kiessling03} for a more recent discussion). 
The presence of the scale $\ell$ naturally `screens' the potential. The Helmholtz equation $\left(\nabla^2-\ell^{-2}\right)\psi_0=4\pi G\rho_0$ sourced by a homogenous density $\rho_0$ (comprised of super-horizon modes) has the well-defined solution $\psi_0 = -4\pi G\rho_0\ell^2 = \,$constant (in space), which has a vanishing force\footnote{Curiously, we see that $\psi_0$ is both proportional to the mean density $\rho_0$ and the surface area of the screened region $4\pi\ell^2$.}.

\section{Conclusions}
\label{sec:conclude}
\noindent
The outputs of large volume cosmological simulations must be interpreted with care in order to correctly account for general relativistic effects at large scales. The standard approach to this problem, as discussed in the Introduction, requires the density contrast $\delta$ of a CDM simulation to be initialised and interpreted as the one in the comoving synchronous gauge, while the velocity field $\mathbf{v}$ and gravitational potential $\psi$ must be interpreted in the conformal Newtonian gauge. In particular, $\psi$ and $\delta$ are then correctly related by Poisson's equation. This rather non-intuitive solution is required in order to explain why the large scale modes of the Newtonian gauge density (which have considerably more power than the synchronous gauge modes at the same scales) should not couple strongly with the small scale modes.

In this paper we developed a simple, physically intuitive recipe to initialise and run an $N$-body simulation \emph{entirely in the conformal Newtonian gauge}, thereby allowing us to directly address and understand this issue. Our main results can be summarised as follows:

\begin{itemize}
\item In the conformal Newtonian gauge, the potential $\psi$ obeys the diffusion \eqn{eq:diffusion}. We showed that this equation can be accurately approximated in the absence of relativistic velocities as the \emph{Helmholtz} \eqn{eq:helmholtz}, with an effective `screening' scale $\ell(a)$ determined by the expansion of the Universe and the linear growth of structure (equations~\ref{eq:elldef} and~\ref{eq:ellapprox}).
\item Since Poisson's equation is replaced by a Helmoltz equation, implementing this screening in an $N$-body simulation requires the trivial change $k^{-2}\to(k^2+\ell^{-2})^{-1}$ in the Green's function kernel relating the potential $\psi$ to the density $\delta$. This also motivates a simple multiplicative correction to convert the density output of a standard simulation into the Newtonian gauge density, as well as a similar approximate gauge transformation for particle positions (equation~\ref{eq:gaugefix} and Figure~\ref{fig:quickfix}).
\item The initial conditions (ICs) for such a `screened' simulation should have $\psi$, $\delta$ and $\mathbf{v}$ initialised in the Newtonian gauge using a modification of the Zel'dovich approximation (Appendix~\ref{app:ICs}). The required transfer functions are standard outputs of Newtonian gauge linear perturbation theory codes.
\item The resulting simulation correctly screens the small scale density from the large scale Newtonian gauge power, and reproduces the linear theory evolution in this gauge at large (including super-horizon) scales (Figure~\ref{fig:Nbody_results}). We note, however, that the particle displacement correction to standard simulations described above is a more accurate (and therefore preferred) way of obtaining the Newtonian gauge density. Running Newtonian gauge simulations thus serves as more of a proof of concept.
\item Although our terminology is borrowed from Debye screening in plasmas, gravity cannot be screened.
We argued that our results are better interpreted as a version of Jeans stability induced by the expansion of the Universe. An upshot of our analysis is the natural resolution of the `Jeans swindle' problem in the presence of super-horizon modes (Section~\ref{sec:analytical}).
\end{itemize}

We end by noting that we have ignored the effects of radiation, baryons and massive neutrinos in our setup, accounting for which is important for precision analyses of large scale structure. As we indicated in Appendix~\ref{app:ICs}, the inclusion of baryons would require modifications of the Zel'dovich approximation even for standard simulations (because the early time velocity fields of baryons and CDM are different); this becomes straightforward in our approach. An obvious next step then is to generalise these modifications to second order Lagrangian perturbation theory which is routinely used in place of the Zel'dovich approximation for generating ICs (see, e.g., \cite{Christopherson+16} for recent results for CDM in the Newtonian gauge). 

Additionally, applying the peak-background split argument \cite{Kaiser84,BBKS86} to the standard and screened simulations, we expect that the halo bias $b^2(k)\equiv P_{\rm h}(k)/P_{\delta}(k)$ should approach the \emph{same constant value} in both types of simulations, at small $k$. It will be very interesting to extend our technique to larger volumes to test this idea. We will return to these issues in future work.

\acknowledgments
\noindent
It is a pleasure to thank T. Padmanabhan for insightful discussions regarding the interpretation of our results. 
We also thank Elisa Chisari for useful correspondence and comments on the draft. We thank Raul Angulo for making the {\sc L-Gadget3} code available to us for the $N$-body simulations discussed in this article. This work was supported by a grant from the Swiss National Supercomputing Centre (CSCS) under project ID s632.

\bibliography{RelScreen}

\appendix
\section{$N$-body results at large scales}
\label{app:interpretNbody}
\noindent
Let us see how the results of standard $N$-body simulations can be interpreted in the context of linear perturbation theory at large scales. For convenience, we set $c=1$ in this and subsequent sections. The argument that follows can be made directly in the language of gauge invariant variables \cite{Mukhanov+92,Haugg+12}, but it will be more convenient for us to start with the metric in the conformal Newtonian gauge \eqref{eq:metric}. In this gauge, the linearised constraint equations $G^\alpha_{\phantom{\mu}0} = (8\pi G)T^\alpha_{\phantom{\mu}0}$, $\alpha=0,..,3$ reduce to \cite{MaBertschinger95}
\begin{align}
\nabla^2\psi_{\rm N} - 3\Cal{H}\left(\p_\tau\psi_{\rm N} + \Cal{H}\psi_{\rm N}\right) 
&=4\pi Ga^2\bar\rho_{\rm m}\delta_{\rm N}\notag\\
\nabla^2\left(\p_\tau\psi_{\rm N} + \Cal{H}\psi_{\rm N}\right) &= -4\pi Ga^2(\bar\rho_{\rm m}+\bar P_{\rm m})\,\theta_{\rm N}
\label{eq:linearisedconstraint}
\end{align}
where the first equation is the same as \eqn{eq:G00=T00},  $\theta\equiv\nabla\cdot\mathbf{v}$ is the velocity divergence and we set $\phi=\psi$. Combining these gives us, in Fourier space,
\be
k^2\psi_{\rm N} = -4\pi Ga^2\bar\rho_{\rm m} \left[\delta_{\rm N} + 3\left(1+w_{\rm m}\right)\frac{\Cal{H}\theta_{\rm N}}{k^2} \right]\,,
\label{eq:linPoisson}
\ee
where $w_{\rm m}\equiv\bar P_{\rm m}/\bar\rho_{\rm m}\approx0$ for CDM. In general one can show, using the gauge transformation between the conformal Newtonian gauge and \emph{any} synchronous gauge \cite{Mukhanov+92,LiddleLyth00}, that
\be
\delta_{\rm N} + 3\left(1+w_{\rm m}\right)\frac{\Cal{H}\theta_{\rm N}}{k^2} = \delta_{\rm syn} + 3\left(1+w_{\rm m}\right)\frac{\Cal{H}\theta_{\rm syn}}{k^2}\,.
\label{eq:NewtVsSyn}
\ee
In particular, for a single fluid comprised of CDM, in the \emph{comoving} synchronous gauge we have $\theta_{\rm syn}=0$, so that \eqn{eq:linPoisson} becomes
\be
k^2\psi_{\rm N} = -4\pi Ga^2\bar\rho_{\rm m}\delta_{\rm syn}\,,
\label{eq:simPoisson}
\ee
where the label `syn' here and below refers to the comoving synchronous gauge. The Newtonian gauge potential therefore obeys Poisson's equation sourced by the density contrast in the comoving synchronous gauge. 

In a standard $N$-body simulation, we have the following chain: the particle velocities (defined as the rate of change of their comoving positions with time) obey \eqn{eq:Euler} in the Eulerian picture; the forces are determined by a potential that obeys Poisson's equation; the density sourcing Poisson's equation is in turn determined by the particle positions. To complete the mapping between simulation and theory, we must then check (a) that the particle positions are consistent with interpreting their density as the comoving synchronous one and (b) that the initial conditions are appropriately set. We discuss these issues in a somewhat broader context in the next two sections, starting with the initial conditions in Appendix~\ref{app:ICs} and then showing, in Appendix~\ref{app:Tmunu}, that particle positions in standard (screened) simulations do correspond to the density contrast in the comoving synchronous (conformal Newtonian) gauge.

\section{Initial conditions}
\label{app:ICs}
The generation of initial conditions (ICs) for a simulation requires some discussion, particularly in the light of our proposed modification in Section~\ref{subsec:helmholtz}. 
Here we focus mainly on the conformal Newtonian gauge, which will bring out some conceptual points that become important when implementing our modification. 
We begin with dark matter and describe the extension to include baryons 
in the ICs at the end. We assume that the initial fluctuations obey Gaussian statistics.

We follow the standard approach in that we use linear perturbation theory to compute the values of $\psi(\mathbf{k},\tau_{\rm i})$, $\delta(\mathbf{k},\tau_{\rm i})$ and $\theta(\mathbf{k},\tau_{\rm i})$ at the initial epoch of the simulation $\tau_{\rm i}$. This means that we can set
\begin{align}
\delta(\mathbf{k},\tau_{\rm i}) &= T_\delta(k,\tau_{\rm i})\,\psi(\mathbf{k},\tau_{\rm i})\notag\\ 
\theta(\mathbf{k},\tau_{\rm i}) &= T_\theta(k,\tau_{\rm i})\,\psi(\mathbf{k},\tau_{\rm i})\,,
\label{eq:deltatheta-init}
\end{align}
where $\psi(\mathbf{k},\tau_{\rm i})$ is a Gaussian random field with zero mean and variance set by the primordial power spectrum $P_\psi(k,\tau_{\rm i}) \propto k^{n_{\rm s}-4}$, and the transfer functions $T_\delta$ and $T_\theta$ are the outputs of the linear theory calculation that satisfy (in the Newtonian gauge) $T_\delta(k,\tau)\to-2$ and $T_\theta(k,\tau)\to 2k^2/(3\Cal{H})$ for $k/\Cal{H}\ll1$ during matter domination, assuming adiabaticity and neglecting decaying modes (see equations~\ref{eq:linearisedconstraint}).

Next, we assign velocities to the simulation particles (which are uniformly distributed on a mesh) using
\be
\mathbf{v}(\mathbf{k},\tau_{\rm i}) = -(i\mathbf{k}/k^2)\,\theta(\mathbf{k},\tau_{\rm i})\,.
\label{eq:v-init}
\ee
In order to reproduce the correct density, however, the particles must be additionally displaced from their mesh locations $\mathbf{q}$ to new locations $\mathbf{x}$ which satisfy
\be
\mathbf{x}(\tau_{\rm i}) = \mathbf{q} + \mathbf{\Psi}(\mathbf{q},\tau_{\rm i})
\label{eq:x-init}
\ee
where the displacement field is given, in Fourier space, by\footnote{This is easily seen by using mass conservation to relate the uniform density of the particles on the mesh to the inhomogenous density of the displaced particles, $\bar\rho(\tau_{\rm i}) = \rho(\tau_{\rm i},\mathbf{x})\,\|\p\mathbf{x}/\p\mathbf{q}\|$, and expanding to lowest order in $\delta$ and the displacement field.}
\be
\mathbf{\Psi}(\mathbf{k},\tau_{\rm i}) = (i\mathbf{k}/k^2)\,\delta(\mathbf{k},\tau_{\rm i})\,.
\label{eq:initdisplacement}
\ee
During matter domination (which is when simulation ICs are typically required) and at subhorizon scales ($k\tau\gg1$), we have the standard result $\delta(\mathbf{k},\tau)=\delta(\mathbf{k})D_1(\tau)=\delta(\mathbf{k})a(\tau)$ if we ignore the decaying mode, and $\theta = -\p_\tau\delta$ \cite{Dodelson03}, which means $\mathbf{v}(\mathbf{k},\tau_{\rm i})=(\p_\tau\ln D_1)\mathbf{\Psi}(\mathbf{k},\tau_{\rm i})$. This is, of course, the Zel'dovich approximation which is usually invoked in setting ICs. Our analysis above has therefore generalised the IC setup in the conformal Newtonian gauge to large scales, showing that the large scale Fourier modes of the particle displacements are no longer proportional to the velocities in this gauge (see, e.g., the super-horizon limits for the transfer functions below equation~\ref{eq:deltatheta-init}). In other words, we must account for corrections to the Zel'dovich approximation in the Newtonian gauge.

What about standard $N$-body simulations which \emph{do} use the Zel'dovich approximation along with a density contrast initialised in the comoving synchronous gauge? As we discussed in Appendix~\ref{app:interpretNbody}, in these simulations one is really interested in $\psi=\psi_{\rm N}$, $\mathbf{v} = \mathbf{v}_{\rm N}$ and $\delta=\delta_{\rm syn}$ at large scales. For these variables, at early times the relation $\mathbf{v}(\mathbf{k},\tau_{\rm i})=(\p_\tau\ln D_1)\mathbf{\Psi}(\mathbf{k},\tau_{\rm i})$ is true at \emph{all} scales, thereby justifying the use of the Zel'dovich approximation.

Finally, if one cares about precision in the initial conditions at the level of a few per cent or less, then the effects of radiation, baryons and massive neutrinos must also be accounted for. For the baryons, for example, this means replacing $T_\delta$ and $T_\theta$ in \eqn{eq:deltatheta-init} with 
\begin{align}
T_\delta &\to f_{\rm c}\,T_{\delta,{\rm c}} + f_{\rm b}\,T_{\delta,{\rm b}}\notag\\
T_\theta &\to f_{\rm c}\,T_{\theta,{\rm c}} + f_{\rm b}\,T_{\theta,{\rm b}}\,,
\label{eq:baryon-deltatheta-init}
\end{align}
where $f_{\rm b}=\Omega_{\rm b}/\Omega_{\rm m}$, $f_{\rm c} = 1-f_{\rm b}$ and the transfer functions $T_{\delta,{\rm c}},T_{\theta,{\rm c}}$ for CDM and $T_{\delta,{\rm b}},T_{\theta,{\rm b}}$ for baryons must be calculated separately in linear theory. 
Notice that, in this case, there is a non-trivial velocity divergence for matter even in the comoving synchronous gauge, since the baryon fluid is not comoving with the dark matter fluid at early times for sub-horizon scales \cite{TseliakhovichHirata10,Angulo+13}. At super-horizon scales, however, the difference between baryons and CDM disappears \cite{MaBertschinger95}.

\section{The energy-momentum tensor of $N$-body simulations}
\label{app:Tmunu}
\noindent
To relate the particle positions in a simulation to a theoretical density contrast in a particular gauge, we need to understand the energy-momentum tensor that is being used by the simulation. In a perturbed FLRW spacetime, the energy-momentum tensor of particles of mass $m$ can be written as an integral over their phase-space distribution function (see, e.g., \cite{BondSzalay83,MaBertschinger95} and references therein):
\be
T^\mu_{\phantom{\mu}\nu} = \frac1{\sqrt{-g}} \int\frac{\der^3P}{(2\pi)^3}\,\frac{P^\mu P_\nu}{P^0}\,f(x^i,P_j,\tau)
\label{eq:Tmunu-fluid}
\ee
where $\der^3P = \der P_1\der P_2\der P_3$, the momentum $P_i$ (the spatial part of the four-momentum, with lower indices) is canonically conjugate to the spatial coordinate $x^i$, and $g$ is the determinant of the metric. The phase-space distribution function $f(x^i,P_j,\tau)$ is a scalar, and the four-momentum $P^\mu$ is constrained by $g_{\mu\nu}P^\mu P^\nu = -m^2$. This expression can be simplified in any specific gauge by introducing the gauge-dependent \emph{proper momentum} $p_i$ measured by an observer at fixed spatial location (with $p^i\equiv\delta^{ij}p_j$), and the corresponding proper energy $E$ which satisfies $E^2=p^2+m^2$, with $p^2=p_ip^i=g_{ij}P^iP^j$. For example, in the conformal Newtonian gauge with metric \eqref{eq:metric}, rewritten for convenience as
\be
\der s^2 = a(\tau)^2\left[-\e{2\phi}\,\der\tau^2 + \e{-2\psi}\,\der\mathbf{x}^2\right]\,,
\label{eq:metric-exp}
\ee
we have $P_i=a\,\e{-\psi}p_i$ and $P^0 = a^{-1}\e{-\phi}E$. The distribution function $f$ is treated as a function of $x^\mu$ and $p_i$, \emph{without} transforming it, so that the particle number is given by $\der N = f\der^3x\der^3P = f\,a^3\e{-3\psi}\der^3x\der^3p$, which is sensible since $a\,\e{-\psi}\der x^i$ is the proper displacement \citep[c.f.][]{MaBertschinger95}. Similar expressions hold in the comoving synchronous gauge.

For a system of CDM particles, the components of $T^\mu_{\phantom{\mu}\nu}$ then become \cite{Bartolo+07}
\begin{align}
T^0_{\phantom{\mu}0} &= -\int\frac{\der^3p}{(2\pi)^3}\,f\,E \equiv -\rho\,,\notag\\
\e{(\phi+\psi)}\,T^0_{\phantom{\mu}i} &= \int\frac{\der^3p}{(2\pi)^3}\,f\,E\,\left(\frac{p_i}{E}\right) \equiv \rho\,v_i\,,\notag\\
T^i_{\phantom{\mu}j} &= \int\frac{\der^3p}{(2\pi)^3}\,f\,E\,\left(\frac{p^ip_j}{E^2}\right) \equiv \rho\,v^i\,v_j\,,
\label{eq:Tmunu-proper}
\end{align}
which leads to the expressions in \eqn{eq:Tmunu} upon neglecting the factor $\e{\phi+\psi}$ for the reasons discussed in the main text. The nice property of the variables $p_i$ and $E$ is that, although they are gauge dependent, the definitions of the energy density $\rho$ and bulk velocity $v_i$ as integrals over $p_i$ have the same form in any gauge \cite{MaBertschinger95}. 

A final simplification occurs upon introducing the \emph{comoving momentum} $Q_i=a\,p_i$, which remains constant in an unperturbed cosmology. Again, we treat the distribution function, without transforming it, as a function of $x^\alpha$ and $Q_i$, and the $T^0_{\phantom{\mu}0}$ component becomes 
\begin{align}
T^0_{\phantom{\mu}0} &= -a^{-3}\int\frac{\der^3Q}{(2\pi)^3}\,f\,E \approx -\frac{m}{a^{3}}\int\frac{\der^3Q}{(2\pi)^3}\,f\,,
\label{eq:T00-comoving}
\end{align}
\emph{in any gauge}, where we used the non-relativistic approximation for the second equality.

We are now in a position to relate these formal expressions to the ``particles'' being tracked in a simulation. The key point to remember is that spacetime in the simulation is \emph{not} perturbed; the simulation simply tracks the evolution of fields in a flat, homogeneous comoving space. The momenta of the simulation ``particles'' are therefore comoving momenta $Q^i = m_{\rm sim}v^i_{\rm sim}$, and the appropriate distribution function is $f\to f_{\rm sim}$ where
\be
f_{\rm sim} = (2\pi)^3\sum_I\,\dir\left(\mathbf{x}-\mathbf{x}_I(\tau)\right)\dir\left(\mathbf{Q}-\mathbf{Q}_I(\tau)\right)\,,
\label{eq:f-sim}
\ee
where $\dir$ is the Dirac delta, the sum is over all the particles in the simulation, and all the effects of $\psi$ are absorbed into the trajectories $\mathbf{x}_I(\tau)$ and momenta $\mathbf{Q}_I(\tau)$. This is sensible, because in the absence of perturbations, the particles would be fixed on the homogenous mesh ($\mathbf{x}_I(\tau)=\mathbf{q}_I$) and their density $\rho_{\rm sim}=-T^0_{\phantom{\mu}0,{\rm sim}}$ would dilute as $\sim a^{-3}$. As it is, performing the momentum integral gives
\begin{align}
T^0_{\phantom{\mu}0,{\rm sim}} &= -\frac{m_{\rm sim}}{a^{3}}\int\frac{\der^3Q}{(2\pi)^3}\,f_{\rm sim}\notag\\
&=  -\frac{m_{\rm sim}}{a^{3}}\,\sum_I\,\dir\left(\mathbf{x}-\mathbf{x}_I(\tau)\right)\notag\\ 
&\equiv -\bar\rho\left(1+\delta_{\rm sim}\right)\,.
\label{eq:T00-sim}
\end{align}
It should be clear from our discussion above that this relation holds in either gauge, Newtonian or synchronous. Since the simulation density is simply a sum over particles in cells, the evolution of $\delta_{\rm sim}$ obeys \eqn{eq:continuity} without the term $3\p_\tau\psi$. The Eulerian velocity field defined using $T^i_{\phantom{\mu}0,{\rm sim}}$ obeys \eqn{eq:Euler} once gravity is included as an external force determined by $\nabla\psi$, and the potential $\psi$ is determined in terms of $\delta_{\rm sim}$.

So, if the particle positions are initialised as in \eqns{eq:x-init} and~\eqref{eq:initdisplacement} using the linear comoving synchronous density $\delta_{\rm syn}$, the velocities initialised using the Zel'dovich approximation, and we use Poisson's equation \eqref{eq:simPoisson} sourced by $\delta_{\rm sim}$ to determine $\psi$, then we are guaranteed to reproduce the linear $\delta_{\rm syn}$ at large scales at all later times. 

On the other hand, if the particle positions are initialised using the conformal Newtonian density $\delta_{\rm N}$, the velocities using the Newtonian gauge velocity divergence $\theta_{\rm N}$, and we use the \emph{Helmholtz} equation \eqref{eq:helmholtz} to calculate $\psi$, then we should reproduce the linear $\delta_{\rm N}$ at large scales at all later times. The results of the main text validate this discussion.

Finally, we note that our use of the simulated particle density directly in the right hand side of the Einstein equation \eqref{eq:G00=T00} may seem at odds with the fact that the Einstein equations require the fully relativistic energy-momentum tensor defined on a perturbed spacetime. Indeed, this is the main reason behind the difference between our prescription of the displacement correction discussed in section~\ref{subsec:gaugefix} and that of \citet{ChisariZaldarriaga11}. Our approach, however, is in the same spirit that allows particle positions of standard simulations to be directly interpreted in the comoving synchronous gauge, which is the usual solution to interpreting large volume simulations as discussed in the Introduction. (The $N$-body gauge defined by \cite{Fidler+15} avoids this complication by explicitly demanding that the particle positions be directly related to the relativistic density contrast.)

What matters at the end, however, is whether or not the variables of interest are evolved correctly by the simulation. The approach of \cite{ChisariZaldarriaga11} is correct because it explicitly uses a relativistic energy-momentum tensor, tracking all terms at the relevant order. In practice, this means that, in addition to the displacement correction (applied at the initial step), the particle density must be corrected by a factor $(1+3\psi)$ (arising from the determinant of the perturbed metric) at every step to convert it into the relativistic Newtonian gauge density. In our approach, on the other hand, the effect of the perturbed metric is absorbed into the time-dependence of the displacement correction \eqref{eq:gaugefix} at all times. This has the added advantage that the initial particle density of a simulation need not be invoked in order to apply the correction. This is also consistent with our prescription for running screened simulations, in which the \emph{initial particle positions} must be set using the relativistic linear theory transfer function in the Newtonian gauge, as discussed in Appendix~\ref{app:ICs}.

\end{document}